\documentclass[twocolumn,showpacs,superscriptaddress,showpacs,floatfix]{revtex4}

\usepackage{epsfig,color}
\usepackage{graphicx}
\usepackage{dcolumn}
\usepackage{bm}
\usepackage{setspace}

\begin{document}

\title{Spin polarization induced tenfold magneto-resistivity of
highly metallic 2D holes in a narrow GaAs quantum well}
\author{X. P. A. Gao}
\email{xuan@cml.harvard.edu}
\affiliation{Harvard University, Cambridge, MA 02138}
\author{G. S. Boebinger}
\affiliation{National High Magnetic Field Laboratory, Florida State University, Tallahassee, FL 32312}
\author{A. P. Mills Jr.}
\affiliation{Physics Department, University of California, Riverside, CA 92521}
\author{A. P. Ramirez}
\author{L. N. Pfeiffer}
\author{K. W. West}
\affiliation{Bell Laboratories, Lucent Technologies, Murray Hill, NJ 07974}

\date{\today}

\begin{abstract}
We observe that an in-plane magnetic field ($B_{||}$) can  induce
an order  of magnitude enhancement  in the  low  temperature ($T$)
resistivity  ($\rho$) of metallic 2D  holes in a narrow (10nm)
GaAs quantum  well. Moreover, we show the first observation of
saturating behavior  of $\rho(B_{||})$ at high $B_{||}$  in GaAs
system, which suggests our large positive $\rho(B_{||})$ is due to
the spin polarization effect alone. We find that this tenfold
increase in $\rho(B_{||})$ even  persists  deeply  into  the  2D
metallic  state  with  the  high $B_{||}$ saturating values of
$\rho$  lower than 0.1h/e$^2$. The dramatic effect of $B_{||}$ we
observe  on  the  highly  conductive  2D  holes (with $B$=0
conductivity as high as 75e$^2$/h) sets strong constraint on
models for the spin dependent transport in dilute metallic 2D
systems.

\end{abstract}
\maketitle

The   metallic behavior   and   metal-insulator   transition (MIT)   in    dilute electrons  or  holes  in  two  dimensional  (2D)
semiconductor  structures have received  much   recent   interest\cite{mitreview,review2004}.  In these low density 2D systems, when the
 carrier density is above the critical density, the system exhibits a significant resistivity drop at low temperature, setting a challenge for
 conventional localization theory.   While novel properties
(e.g. the dramatic change in compressibility at
MIT\cite{jiangcompress}, the  anomalous   thermopower
\cite{thermopower} and enhanced  phonon coupling
\cite{phononcoupling}  effects) are continuing  to  be discovered
in  this 2D   metallic state,  many   critical  issues  still
remain unresolved.  Outstanding   questions   include: Does the
Fermi liquid (FL) phenomenology still hold for the 2D metallic
state where $r_s\gg$1? Is this MIT a true quantum phase transition
or simply a   crossover at finite temperature?  And most
importantly, what is the mechanism for the resistivity drop?

The spin degeneracy is believed to be essential for inducing the
metallic resistivity, as it was   found that an in-plane magnetic
field $B_{||}$ suppresses the metallicity and    in    some cases
drives    the   system
insulating\cite{simonian,yoon,papadakis,Gao,tutucSDHBp}. Recent
experiments on GaAs quantum well (QW) further revealed an
intriguing $B_{||}$ insensitivity of the energy scale of the 2D
metal as well as the FL-like logarithmically diverging $\rho(T)$
of 2D holes in strong $B_{||}$\cite{Gao}. Many theoretical models
were proposed to explain the $B_{||}$ destruction of the 2D
metallic transport, such as the superconductivity
scenario\cite{Philips}, the FL-Wigner solid coexisting
microemulsion model\cite{spivakkivelson}, or screening model based
on conventional FL wisdom \cite{goldJETP,dassarmaparaB}. It was
even noticed that positive $\rho(B_{||})$ can be induced by the
magneto-orbital effect of $B_{||}$ due to the finite thickness of
the sample, without involving any spin effect\cite{dassarmaB||}.

In this paper, we   present a study of   the in-plane magnetic
field   induced magneto-transport of a low density 2D hole system
(2DHS) in a {\it narrow} (10nm wide) GaAs  QW down to as low as
$T$=20mK. We show that the resistivity   of our 2DHS can increase
by nearly an order of magnitude followed by a saturation as
$B_{||}$  increases, similar to  the case  for  Si-MOSFET's.  In
contrast to previous experiments on GaAs heterostructures or wider
QWs \cite{yoon,papadakis,tutucSDHBp,zhunGaAschi}, our result
clearly disentangles         the spin effect from the orbital
effect\cite{dassarmaB||} in the $B_{||}$-dependent transport
studies of the 2D metallic state.  Moreover, it is striking that
this spin polarization induced tenfold magneto-resistivity even
persists deeply into the metallic state where the conductivity
$\sigma$  is as high as   75e$^2$/h at $B$=0. In-plane
magneto-transport has been extensively calculated for low density
2D systems within the screening theory for FL. For weak disorder,
semi-classical calculations based on $T$- and $B$-dependent
screening\cite{dassarmaparaB} showed good agreement with highly
conductive Si-MOSFET's, in which a factor 3 to 4 increase in
$\rho(B_{||})$ and a weak $T$-dependent $\rho(T)$ in the spin
polarized state were observed\cite{tsui,shashkin05}. Refined
screening models including exchange and correlation effects may
produce a larger increase in $\rho(B_{||})$, but only when
disorder is sufficiently strong and carrier density sufficiently
low to be in the vicinity of the MIT\cite{goldJETPdisorder}.   Our
observation of such large $\rho(B_{||})$ for metallic 2DHS with
$\sigma\gg e^2/h$(or $k_Fl\gg$1) calls for further theoretical
understanding of spin-dependent transport in dilute metallic 2D
systems with strong correlations and weak disorder.

Our experiments were performed on a high mobility low-density 2DHS
in a  10nm wide GaAs QW similar to
ref\cite{Gao,GaoHall,phononcoupling}. The sample was grown on a
(311)A GaAs wafer using Al$_{.1}$Ga$_{.9}$As barrier. Delta-doping
layers of Si dopants were symmetrically placed above and below the
pure GaAs QW. Diffused In(1$\%$Zn) was used as contacts. The hole
density $p$ was   tuned    by a backgate voltage. The ungated
sample has a low temperature hole mobility, $\mu\approx 5\times
10^5$cm$^2$/Vs, and  a density $\sim$1.6$\times$10$^{10}$cm$^{-2}$
from   doping.  The sample was prepared in the   form  of   Hall
bar,  with   an approximate total   sample area  0.2cm$^2$. With
the relatively large sample area and the  measuring current
induced heating power at the level of fWatts/cm$^2$, the   low
density 2DHS  can   be reliably cooled down   to  20mK with
negligible self-heating\cite{phononcoupling}. All the data in this
paper were taken with the current along the [$\underline 2$33]
high mobility direction. $B_{||}$ was also applied along the
[$\underline 2$33] direction, where the effective
g-factor$\sim$0.6 \cite{Gao}. During the experiments, the sample
was immersed in the $^3$He/$^4$He mixture in a top-loading
dilution refrigerator.

\begin{figure}[btph]
\centerline{\psfig{file=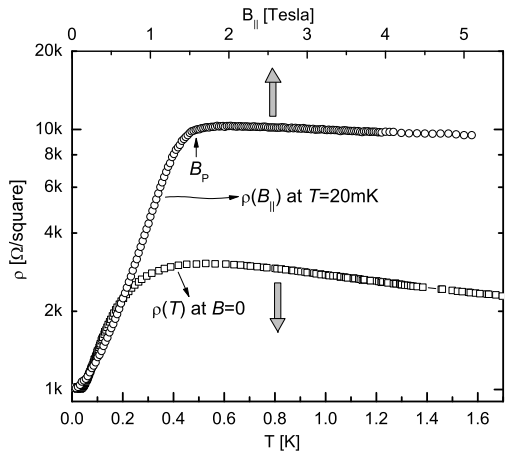,width=7.5cm}}
\caption{Resistivity vs.  $T$ at zero magnetic field and
resistivity vs in-plane magnetic field $B_{||}$ at $T$=20mK for a
2DHS in 10nm wide GaAs QW. The hole density
$p$=1.35$\times$10$^{10}$cm$^{-2}$. Similar to Si-MOSFET's,
$\rho(B_{||})$ shows saturation above $B_P$=1.5T, in contrast to
wider GaAs QW's or
heterostructures\cite{yoon,papadakis,tutucSDHBp,zhunGaAschi}.}
\label{fig1}
\end{figure}

Figure \ref{fig1}  shows  the  $T$=20mK $\rho$ vs. $B_{||}$ and
the zero magnetic field $\rho$ vs. $T$ data of our 2DHS with
$p$=1.35$\times$10$^{10}$cm$^{-2}$ in the metallic phase of MIT.
At $B$=0, $\rho$ shows a factor of three drop below 0.4K. The
$\rho(B_{||})$ curve shows a very large magneto-resistivity below
1.5T and a nearly constant $\rho$ at higher $B_{||}$. This
behavior is rather similar to the $\rho(B_{||})$  data in
Si-MOSFET's and the magnetic field $B_P$ at which $\rho$ starts
saturating was identified to be the field when the system obtains
full spin polarization\cite{okamotoSDHBp, vitkalov, kravchenkoBp}. We mention
that all previous $\rho(B_{||})$ data on GaAs 2D electron/hole
systems show somewhat different behavior: the resistivity
continuously increases with a reflection point around $B_P$ upon
applying $B_{||}$ \cite{yoon,papadakis,tutucSDHBp,zhunGaAschi}. We
believe that the saturating behavior of our $\rho(B_{||})$ here at
$B_{||}>B_P$ is due to the smaller thickness of our QW. The
constant $\rho(B_{||})$ above $B_P$ of our QW also suggests that
the magneto-orbital effect related scattering\cite{dassarmaB||} is
small in our case. For GaAs heterostructures or wider QW's (and
low carrier concentration), the magnetic length at several Tesla
becomes comparable or smaller than the width of the 2D
electron/hole wavefunction in the $z$-direction and the
magneto-orbital effect can induce a continuous positive
magneto-resistivity as discussed by Das Sarma   and
Hwang\cite{dassarmaB||}. Note that for experiments on Si-MOSFET's
the confinement in the $z$-direction is also narrow and a
saturation in $\rho$ is often observed after an increasing $\rho$
at low $B_{||}$\cite{okamotoSDHBp,vitkalov,kravchenkoBp}. Thus our data
suggest that the thickness effect is certainly able to explain
most of the differences in $\rho(B_{||})$ behavior between GaAs
and Si-MOSFET systems, although the valley degeneracy may play
some additional role.

\begin{figure}[htbp]
\centerline{\psfig{file=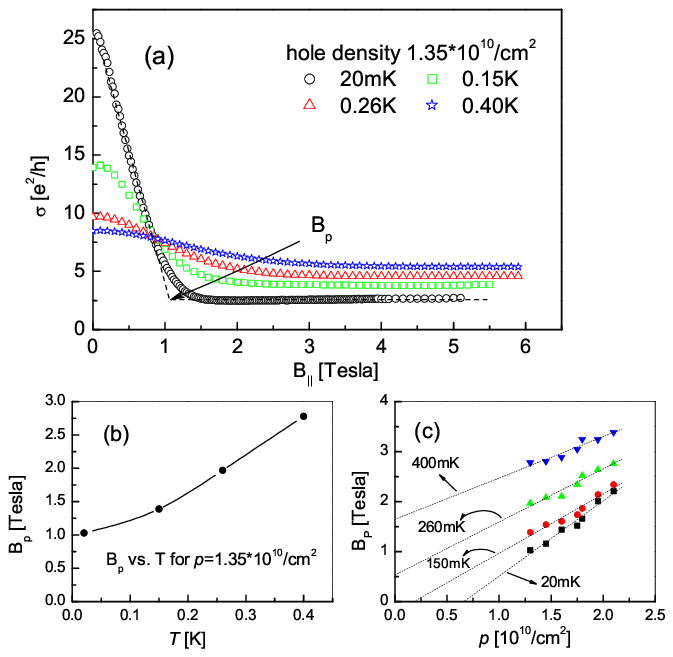,width=8.5cm}}
\caption{(color online)(a)   The   conductivity    $\sigma$   of
2DHS  with $p$=1.35$\times$10$^{10}$cm$^{ -2}$  as  a  function of
the in-plane  magnetic field  $B_{||}$  at 20mK,  0.15K, 0.26K and
0.40K.  The  magnetic field  $B_P$ above   which  all   the spins
are  polarized   is    determined   by     the intersection   of
linear   extrapolations of $\sigma(B_{||})$ at low  and  high
field  regions. (b) The temperature   dependence of $B_P$, with
the  dashed line as a guide to the  eye. (c)$B_P$ as a  function
of hole density $p$  at  20mK, 0.15K,  0.26K and 0.4K. The dotted
lines are the  linear fittings  to the  data. It can be    seen
that finite  temperature strongly  affects   the behavior of
$B_P(p)$,  and    it is   only  at   the lowest temperatures
$B_P(p)$  linearly extrapolates to zero at a finite density.
}\label{fig2}
\end{figure}
Now we discuss how the temperature affects the magneto-transport.
In Fig.\ref{fig2}a we  plot the $\sigma(B_{||})$ for
$p$=1.35$\times$10$^{10}$cm$^{-2}$ at  20mK,  0.15K, 0.26K  and
0.40K. All the iso-thermal $\sigma(B_{||})$  curves  cross  around
1.2T,  indicating  the  `$B_{||}$ induced
MIT'\cite{yoon,papadakis,Gao}.  As suggested by Vitkalov {\it et
al.}\cite{vitkalovdisorder}, we can determine the magnetic field
$B_P$ for the onset of full spin polarization of $delocalized$
holes by the intersection of linear extrapolations of
$\sigma(B_{||})$ at low and high field regions.  Nonetheless, we
obtain a $B_P$ only 10$\%$ higher if the extrapolating  process is
applied to $\rho(B_{||})$, suggesting most holes are delocalized.
We find that $B_P$ is strongly temperature dependent.    As  one
can see in Fig.\ref{fig2}b where  $B_P(T)$ is plotted  for this
density, $B_P$ at $T$=20mK is only   40$\%$ of  its    value  at
0.4K. Since   the   $B_P$ is generally   regarded as the magnetic
field required to  fully polarize   the spins  of  delocalized
carriers\cite{okamotoSDHBp,vitkalov}, one natural interpretation of the
$T$-dependent $B_P$ is that {\it the spin susceptibility
$\chi$ is largely enhanced as $T$ is reduced}. This strong
$T$-dependent $B_P$ has implications in other models as well. For
instance, in the `microemulsion' model it would mean that it
requires much less Zeeman energy to solidify the FL phase at lower
temperatures\cite{spivakkivelson}.

Figure\ref{fig2}c shows the density dependence  of $B_P$ at 20mK,
0.15K, 0.26K and 0.4K. Previously, extrapolating $B_P(p)$ to
$B_P$=0 was  used as a way to  determine if a  ferromagnetic
instability exists  in the
system\cite{review2004,zhunGaAschi,kravchenkoBp}.  If $B_P(p)$
extrapolates  to zero at   a finite   density,  then   such
density   corresponds  to  the ferromagnetic  instability. It can
be  seen  in    our Fig.\ref{fig2}c that our $B_P(p)$ data taken
at different $T$ extrapolate to zero at different densities. Only
at low temperatures $B_P(p)$ linearly extrapolates to  zero at  a
finite  density. At $T$=20mK, $B_P(p)$ extrapolates to zero at a
density very close to the critical density of the $B=0$
MIT\cite{footnote}. Our $T$-dependent study of $B_P(p)$ is
consistent with Si-MOSFET's\cite{sarachikchiT}. Although the
meaning of a diminishing $B_P$ at finite $p$ is controversial and
may actually be associated with other physics (e.g. the
instability to crystallization instead of
ferromagnetism)\cite{spivakkivelson,spivakkivelsonBp}, our
experiment on p-GaAs corroborates the universal existence of this
behavior.

\begin{figure}[hbtp]
\centerline{\psfig{file=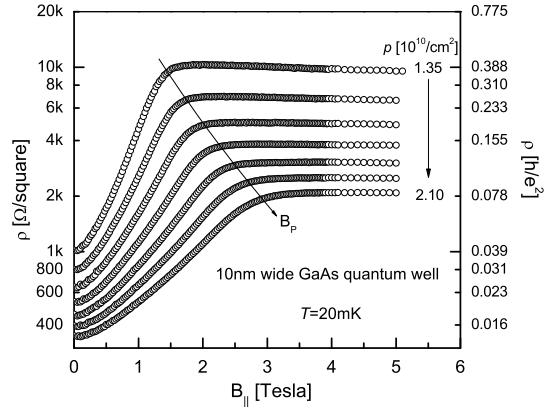,width=8cm}}
\caption{Resistivity $\rho$ vs. $B_{||}$ at $T$=20mK of 2D holes
in a 10nm wide  GaAs quantum well.  The hole densities are 1.35,
1.48, 1.60, 1.73, 1.85, 1.98 and 2.10$\times$10$^{10}$cm$^{-2}$
from top to bottom. The arrow marks the positions of $B_P$, the
magnetic field above which $\rho(B_{||})$ shows saturation. Note
that the almost factor of 10 increase in $\rho$ persists deeply
into the metallic phase with high $B_{||}$ values of
$\rho<0.1$h/e$^2$.} \label{fig3}
\end{figure}
\begin{figure}[htbp]
\centerline{\psfig{file=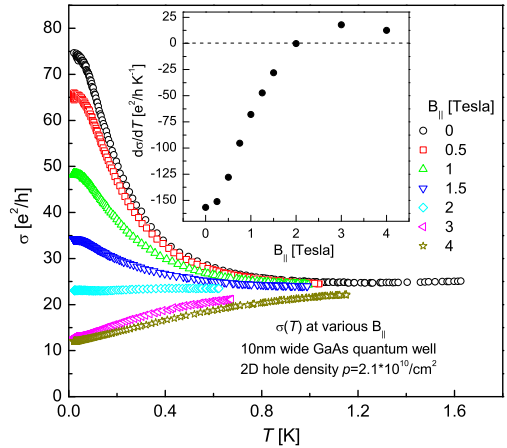,width=8cm}} \caption{(color
online) 2D  hole conductivity  $\sigma$ vs.  $T$ at  various  in
-plane      magnetic    field     $B_{||}$'s.    The      density
$p$      is 2.1$\times$10$^{10}$cm$^{-2}$. The  inset shows  the
slope of   $\sigma(T)$  as a  function  of $B_{||}$.  the  slope
of $\sigma(T)$ is obtained by fitting data linearly between 0.06K
and 0.2K.} \label{fig4}
\end{figure}

Figure\ref{fig1} has shown that the narrow p-GaAs QW responds to
$B_{||}$ quite similarly to Si-MOSFET's, where $\rho(B_{||})$
shows large increase at low $B_{||}$ and a saturation at high
field. It is striking that for our p-GaAs, this order of magnitude
positive $\rho(B_{||})$ even persists deeply into the metallic
phase where $\rho\ll$h/e$^2$ for both the $B_{||}$=0 low spin
polarization phase and the high $B_{||}$($>B_P$) spin-polarized
phase\cite{spinnote}. Fig.\ref{fig3} shows $\rho(B_{||})$ at 20mK
for seven densities up to 2.1$\times$10$^{10}$cm$^{-2}$. The large
enhancement and saturation in $\rho(B_{||})$ are observed for all
these densities. Note that for $p$=2.1 (the lowest curve) the high
field ($B>B_P$) value of $\rho$ is clearly below 0.1h/e$^2$. {\it
For Si-MOSFET system with comparable resistivity, $\rho$ usually
shows only a factor of 3-4 increase below
$B_P$}\cite{kravchenkoBp,vitkalov,goldJETPdisorder}, in agreement
with the screening
model\cite{goldJETP,dassarmaparaB,goldJETPdisorder}. Note that the
screening model predicts at most a factor of four increase in
$\rho$ due to reduced screening from the lifted spin degeneracy
for $\rho\ll h/e^2$ \cite{goldJETP,dassarmaparaB}. Only very near
the critical density of the MIT can the Si-MOSFET show
$\rho(B_{||})/\rho(0)>$4, resulting from many-body and strong
disorder effects in the screening model\cite{goldJETPdisorder}.
Moreover, the original publication of screening theory predicts a
weak metallic like $\sigma(T)$ at
$B_{||}>B_P$\cite{dassarmaparaB}, in disagreement with our data in
Fig.4 below. It is possible that exchange (Fock) term of the
electron-electron interaction\cite{ZNA} could account for the
difference; however, to date, the only FL theory including both
Hartree and Fock interactions\cite{ZNA}, is perturbative, valid
only at $T\ll T_F$ and not applicable to our experimental regime.
More sophisticated non-perturbative Fermi liquid calculations are
needed for further comparison with our data.

The dramatic effect of spin polarization induced by $B_{||}$ on
our dilute 2DHS also exhibits in the temperature dependence of the
conductivity. In Fig.\ref{fig4} we plot $\sigma(T)$ at various
$B_{||}$ for $p$=2.1$\times$10$^{10}$cm$^{-2}$. At $B$=0,the 2DHS
shows a factor  of three increase in   the conductivity below 0.8K
and the low $T$  conductivity is as high  as 75e$^2$/h. With the
application of $B_{||}$, the metallic conductivity enhancement
becomes smaller and eventually $\sigma(T)$ turns into
insulating-like(d$\sigma$/d$T>$0) above 2T. In the inset of
Fig.\ref{fig4} we plot d$\sigma$/d$T$, the slope of $\sigma(T)$,
as a function of $B_{||}$ to demonstrate this strong effect of
$B_{||}$ on the 2D metallic transport. It can be seen that the
absolute values of the slope of $\sigma(T)$ differ  by about  a
factor  of ten  between the  zero and high field regimes. A
similar effect was also seen in Si-MOSFET's\cite{tsui}.

The $B_{||}$  suppression of 2D metallic transport  was attributed
to the FL interaction correction effects in the ballistic
regime\cite{ZNA} in various recent experimental
papers\cite{proskuzna}. Here we do not attempt to fit our data to
extract the FL parameter $F_0^\sigma$ since we believe that the
$perturbative$ FL calculation should not be taken as a
quantitative theory for our {\it order of magnitude} increase in
$\rho(B_{||})$.  Recent Hall coefficient measurements on similar
samples also provide experimental evidence against the interaction
correction interpretation for the metallic $\sigma(T)$ at
$B$=0\cite{GaoHall}, further reflecting the fact that the $T\ll
T_F$ theory is inapplicable to our data\cite{dassarmahall}. A
non-perturbative FL calculation including both the Hartree and
Fock interaction terms and extending to temperatures $T>T_F$ would
be required to make a direct comparison with our data. Another
possible explanation for our large magnetoresistivity effect comes
from a non-perturbative non-FL approach: it has been theoretically
argued that intermediate phases (`microemulsions') exist in clean
2D systems between the FL phase and the Wigner solid
phase\cite{spivakkivelson,spivakkivelsonBp}. In such a scenario,
the dramatic suppression of the slope of $\sigma(T)$ by an
in-plane magnetic field $B_{||}$ would be analogous to the
magnetic field effect on the Pomaranchuk effect in
$^3$He\cite{spivakkivelson}. It will be of interest to develop
more quantitative calculations based on such model for a direct
comparison with our experimental data.

The authors are indebted to S.A.  Kivelson  and  B.  Spivak  for
discussions and encouragement which stimulated the present paper.
We also thank S. Das Sarma for bringing ref.\cite{dassarmaparaB}
to our attention. The NHMFL is supported by the NSF and the State
of Florida.


\begin{references}

\bibitem{mitreview}E. Abrahams,  S.V. Kravchenko,  and M.P.  Sarachik, {\it Rev.
Mod. Phys.} {\bf 73}, 251 (2001).

\bibitem{review2004}S. V.  Kravchenko, M.  P. Sarachik,  {\it Rep.  Prog. Phys.}
{\bf 67}, 1 (2004).

\bibitem{jiangcompress}S. C. Dultz and H. W. Jiang, {\it Phys. Rev. Lett.}
{\bf 84}, 4689 (2000).

\bibitem{thermopower}L. Moldovan {\it et al.}, {\it Phys. Rev. Lett.} {\bf 85},
4369 (2000).

\bibitem{phononcoupling}X. P.A. Gao {\it et  al.}, {\it Phys.  Rev. Lett.} {\bf 94}, 086402 (2005).

\bibitem{simonian}D. Simonian, S.V. Kravchenko, M.P. Sarachik, and V.M. Pudalov,
{\it  Phys. Rev.  Lett.} {\bf  79}, 2304  (1997).

\bibitem{yoon}J. Yoon,  C.C. Li,  D. Shahar,  D.C. Tsui,  and M.  Shayegan, {\it
Phys. Rev.  Lett.} {\bf  84}, 4421  (2000).

\bibitem{papadakis} S. J. Papadakis, E. P. De Poortere, M. Shayegan, and R. Winkler, {\it Phys. Rev. Lett.} {\bf 84}, 5592 (2000).

\bibitem{Gao}X. P.A. Gao, A.P. Mills, Jr., A.P. Ramirez, L.N. Pfeiffer, and K.W.
West, (a) {\it  Phys. Rev.  Lett.} {\bf  88}, 166803 (2002) ; (b) {\it  ibid} {\bf  89}, 016801  (2002).

\bibitem{tutucSDHBp}E. Tutuc, E.P. De Poortere, S.J. Papadakis, and M.  Shayegan
{\it  Phys. Rev.  Lett.}  {\bf  86}, 2858  (2001);E.  Tutuc, S. Melinte, and  M.
Shayegan, {\it Phys. Rev. Lett.} {\bf 88}, 036805 (2002).


\bibitem{Philips} P. Phillips, Y. Wan, I. Martin, S. Knysh, D. Dalidovich, {\it Nature} {\bf 395}, 253 (1998).

\bibitem{spivakkivelson}B. Spivak, {\it Phys.  Rev. B} {\bf 67},  125205 (2002);
B. Spivak and S.  A. Kivelson, {\it ibid.}  {\bf 70}, 155114 (2004);Reza  Jamei,
Steven Kivelson, and Boris Spivak, {\it Phys. Rev. Lett.} {\bf 94}, 056805 (2005).

\bibitem{goldJETP}V. T. Dolgopolov and A. Gold, {\it JETP Lett.} {\bf 71}, 27 (2000).

\bibitem{dassarmaparaB}S. Das Sarma and E.H. Hwang, {\it Phys. Rev. B}
 {\bf 72}, 035311 (2005); {\bf 71}, 195322 (2005); {\bf 72}, 205303 (2005).

\bibitem{dassarmaB||} S. Das Sarma and E.  H. Hwang {\it Phys. Rev. Lett.}  {\bf
84}, 5596 (2000).

\bibitem{zhunGaAschi}J. Zhu, H. L. Stormer,  L. N. Pfeiffer, K. W.  Baldwin, and
K. W. West, {\it Phys. Rev. Lett.} {\bf 90}, 056805 (2003).

\bibitem{tsui}Yeekin Tsui, S. A. Vitkalov, M. P. Sarachik, and T. M. Klapwijk,
{\it Phys. Rev. B} {\bf 71}, 113308 (2005).

\bibitem{shashkin05}A.A. Shashkin {\it et al.}, {\it Phys. Rev. B} {\bf 73}, 115420 (2006).

\bibitem{goldJETPdisorder} A. Gold, {\it JETP Lett.} {\bf 72}, 274 (2000).

\bibitem{GaoHall}X. P. A.  Gao {\it et  al.}, {\it Phys.  Rev. Lett.} {\bf  93},
256402 (2004).


\bibitem{okamotoSDHBp}T. Okamoto, K. Hosaya, S. Kawaji, and A. Yagi, {\it  Phys.
Rev.  Lett.} {\bf  82}, 3875  (1999).

\bibitem{vitkalov} Vitkalov {\it et al., Phys. Rev. Lett.} {\bf 85}, 2164 (2000).

\bibitem{kravchenkoBp}A. A. Shashkin, S. V. Kravchenko, V. T. Dolgopolov, and T.
M. Klapwijk {\it Phys. Rev. Lett.} {\bf 87}, 086801 (2001).

\bibitem{vitkalovdisorder}S. A. Vitkalov, M. P. Sarachik, and T. M. Klapwijk,
{\it Phys. Rev. B} {\bf 65}, 201106 (2002).


\bibitem{footnote}The $\sigma(T)$ data for a sample similar to  the one in this
paper can be  found in Fig.1 of ref.\cite{GaoHall}.

\bibitem{sarachikchiT} M.P. Sarachik and S.A. Vitkalov, cond-mat/0209113; also
J. Phys. Soc. Jpn. Suppl. A 72, 57-62 (2003).

\bibitem{spivakkivelsonBp} B. Spivak and S.A. Kivelson, cond-mat/0510422; {\it Journal De Physique IV} {\bf 131}, 255 (2005).

\bibitem{spinnote}Low $T$ Shubnikov de-Haas oscillation data
suggest that there is significant spin splitting in our sample
even at $B$=0, which increases from 20$\%$ to 32$\%$ as the
density decreases from 2.35 to 1.35$\times$10$^{10}$cm$^{-2}$
\cite{GaoHall}.

\bibitem{ZNA}G. Zala, B. N. Narozhny, and I. L. Aleiner, (a) {\it Phys. Rev. B}  {\bf64}, 214204 (2001); (b) {\it ibid.}, {\bf 65}, 020201 (2002).

\bibitem{proskuzna}Y. Y. Proskuryakov {\it et al.}, {\it Phys. Rev. Lett.}  {\bf 89}, 076406 (2002);V. M. Pudalov {\it et al.}, {\it ibid} {\bf 91},
126403 (2003);S. A. Vitkalov, K. James,  B. N. Narozhny, M. P.  Sarachik,
and T. M. Klapwijk, {\it Phys. Rev. B} {\bf 67}, 113310 (2003).

\bibitem{dassarmahall}S. Das Sarma and E. H. Hwang, {\it Phys. Rev. Lett.}
{\bf 95}, 016401 (2005).

\end{references}
\end{document}